\documentclass[prl,twocolumn,showpacs,a4paper]{revtex4}
\usepackage{graphicx}%
\usepackage{dcolumn}
\usepackage{latexsym}

\def\mis{$\mu ms^{-1}$}
\def\mum{$\rm{\mu m}$}

\makeatletter
\def\btt#1{\texttt{\@backslashchar#1}}%
\DeclareRobustCommand\bblash{\btt{\@backslashchar}}%
\makeatother

\begin{document}
\preprint{HEP/123-qed}

\title[ ]%Short Title
{Hexagonal eutectic solidification patterns operating near a marginal stability point }
\author{$\rm{Mika\ddot{e}l}$  Perrut}
\altaffiliation{Present address: Laboratoire L\'eon Brillouin, CEA-CNRS, CE-Saclay, F-91191 Gif-sur-Yvette, France}
\author{Silv\`ere Akamatsu}
\author{Sabine Bottin-Rousseau}
\email{bottin@insp.jussieu.fr}
\author{Gabriel Faivre}

% \thanks{Also at Physics Department, XYZ University.}%Lines break
 % automatically or can be forced with \\  
\affiliation{Institut des Nanosciences de Paris, CNRS UMR 7588, 
Universit\'e  Pierre-et-Marie-Curie, Campus
Boucicaut, 140 rue de Lourmel, 75015 Paris,  France} 
\date{\today}% It is always \today, today, but you may specify any
             % date with \date.  
 
\begin{abstract}
We study the long-time dynamics of hexagonal directional-solidification patterns in bulk samples of a transparent eutectic alloy using an optical method which permits real-time observation of the growth front. A slow dilatation of the patterns due to a slight curvature of the isotherms drives the system into a permanent regime, close to the threshold for the rod splitting instability. Thus an apparently minor instrumental imperfection suffices to maintain the system near a marginal stability point. This answers the long-standing question of spacing selection in bulk eutectic growth.
\end{abstract}

%Valid PACS numbers may be entered using the \verb+\pacs{#1}+ command.

% PACS, the Physics and Astronomy Classification Scheme. 
\maketitle

%\tableofcontents
%\pacs{PACS Numbers :} %46.30.Lx, 62.20.Dc, 68.55.Jk}

Directional solidification of nonfaceted binary eutectics ({\it{i.e.}} binary alloys, which are two-phased in the solid state, both solid phases displaying nonfaceted melt growth) gives rise to diffusion-controlled out-of-equilibrium patterns consisting of more or less periodic arrangements of the two solid phases. These patterns have spacing values $\lambda$ lying usually in the 1-10\mum~range for a solidification rate V of 1\mis, and thus contain a large number of repeat units in bulk samples. What is their behavior at long solidification times  --in other words, what are their permanent regimes, if any, at given control parameters (V, and the alloy composition C)-- is a long-standing open question \cite{JackHunt66, Langer1, TriMas91, Caroli94}. Discussions have focussed on two apparently unrelated sets of results. On the one hand, theory \cite{JackHunt66, Karma96} supported by experiments  in thin samples of transparent eutectics \cite{Ginibre97,akaplapp02} indicates that the permanent regime  of an ideal system must be a  perfectly periodic pattern, the spacing of which depends on initial conditions. On the other hand, numerous experiments in bulk metallic eutectics displayed permanent regimes characterized by a broad spacing distribution independent of initial conditions. The mean value $\bar{\lambda}$ of this distribution varies with $V$ approximately as  $V^{-0.5}$  \cite{ TriMas91,chapeau,Drevet96,Ratke00}. This gap between theory and experiments is called the spacing-selection problem in bulk eutectic growth.

To advance this, and other, fundamental problems in eutectic pattern formation, we built an optical device, which permits real-time observation of transparent eutectic growth fronts in bulk samples. A detailed presentation of this device can be found elsewhere  \cite{setup07}. In brief, we use glass crucibles made of two  glass plates  separated by 0.4mm-thick plastic spacers. The cross-section of the inner volume, and thus the area occupied by growth front, is a $0.4\times6$mm${^2}$  rectangle. The crucibles are filled with a  transparent eutectic alloy, and placed between a cold oven and a hot oven. The thermal gradient ($G=8\pm 1~\rm{Kmm^{-1}}$) is established by thermal diffusion along the sample. The samples are pulled at a rate $V$ toward the cold side of the thermal gradient. The growth front lies in the 5-mm wide gap between the ovens and is observed through the liquid and a sample wall with a long-distance microscope. The microscope and the direction of lighting are tilted in the plane $yz$,  where  {\bf z} is the  growth axis and  {\bf y} is the normal  to the sample walls, and make angles of about  $40\rm{^o}$ with  {\bf z}. The resulting one-directional compression of the image  is corrected numerically. To visualize the growth front itself (not the underlying solid), a dark-field image is formed with the light emerging from the interfaces between the liquid and one of the eutectic solid phases.  Two additional refinements in the preparation of the experiments are important. Single eutectic grains are grown  to eliminate the perturbations that are generated by the anisotropy of the surface tensions, in particular, near eutectic grain boundaries  \cite{EutGrains2}.  The thermal bias $G_y$, {\it{i.e.}} the (not wanted) $y$-component of the thermal gradient, is  made as small as possible in the region of the growth front  using a device included in the ovens for this purpose.

\begin{figure}[htbp]
\centering 
{\includegraphics*[width=8cm]{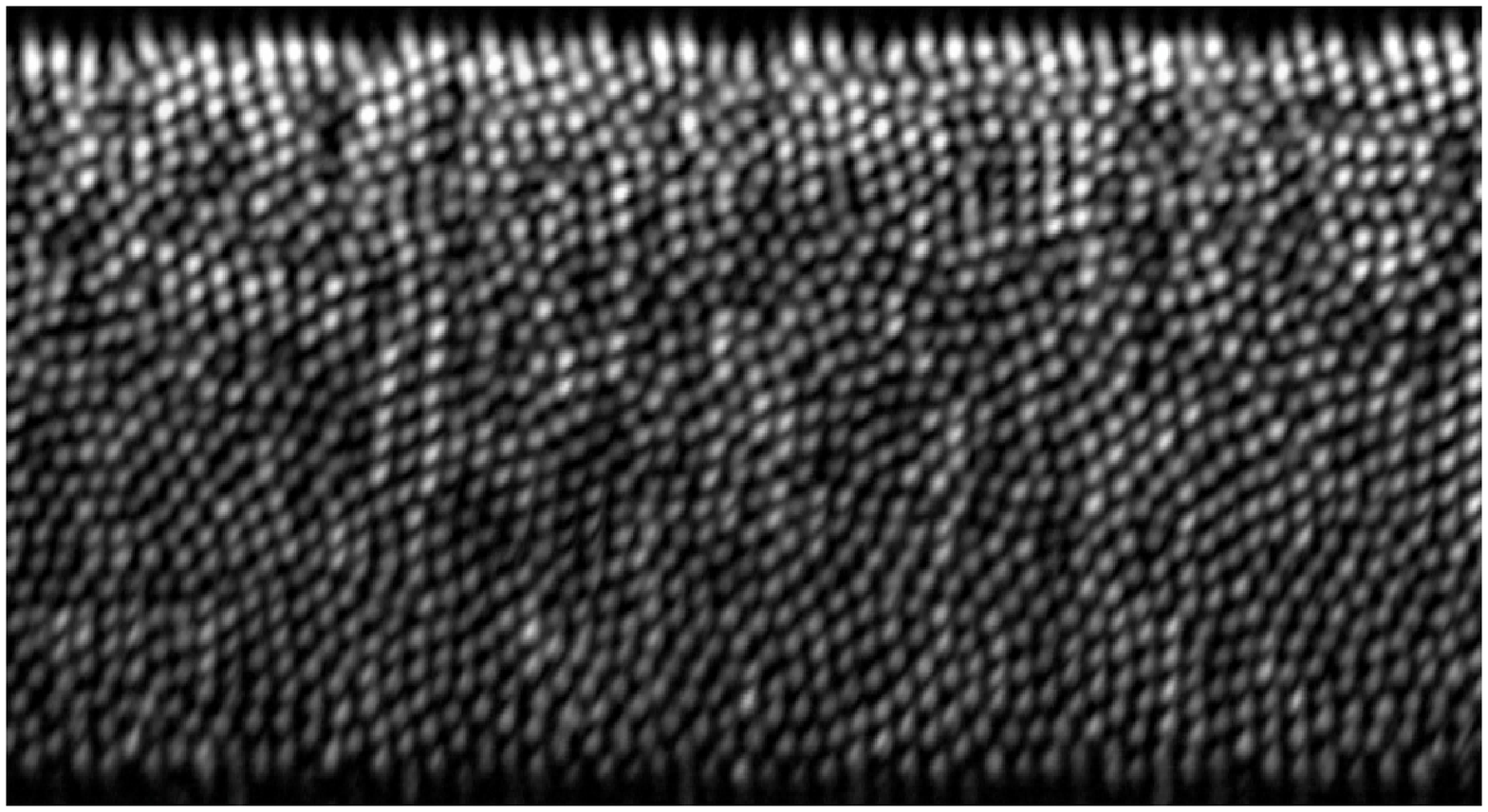}}
{\includegraphics*[width=8cm]{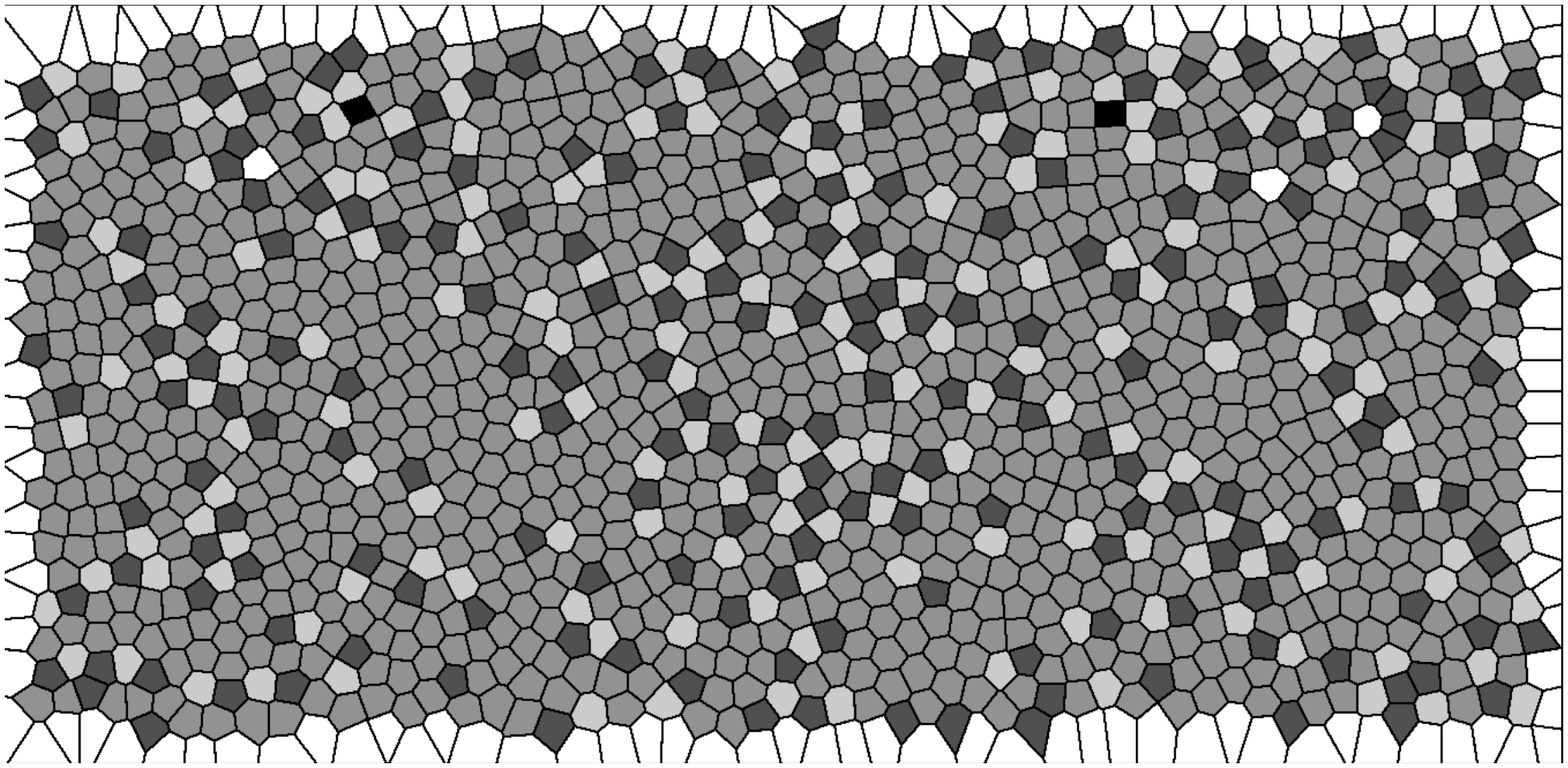}}
\caption{a) Dark-field image of a growth front  of a eutectic SCN-DC alloy directionally solidified  at $V=0.035 \mu ms^{-1} $. The growth direction  {\bf z} is pointing towards the reader. The normal  {\bf y} to the sample walls is pointing upwards. The vertical side of the field of view coincides with the $400\mu m$-long width of the growth front. b) Voronoi  diagram of the growth pattern.  Cells have been greyed according their nearest-neighbor numbers (medium grey: 6-fold coordinated cells). }
\label{manip1}
\end{figure}

This method of observation was successfully used during a study of lamellar (banded) eutectic growth patterns \cite{akabottin04}. In this Letter, we employ it to study  the  spacing-selection problem in the rod-like (hexagonal) growth patterns of the alloy succinonitrile-(D)camphor (SCN-DC) at eutectic concentration (14 mol\% of DC). The basic properties of this alloy were studied experimentally \cite{scndc}. The scaling constant for eutectic growth  ($\lambda_m^2V=10.2\pm1.5 \mu \rm{m^3s^{-1}}$, where  $\lambda_m$ is the minimum-undercooling spacing \cite{JackHunt66}) was determined, and it was shown that anisotropy effects are particularly weak in this alloy. The eutectic solid phases are almost pure cubic SCN and hexagonal DC, respectively. The envelope of the growth front nearly follows the isotherm corresponding to the eutectic temperature of the system ($T_E =  38.3 \rm{^oC}$). The growth patterns are arrays of disks of DC-liquid interfaces  ("DC caps") embedded in a continuous SCN-liquid interface. We selected the light emerging from the DC caps to form the images of the growth patterns, which thus appeared as arrays of bright spots on a dark background (Fig. \ref{manip1}). These spots correspond to the caustics of the light rays transmitted through the DC rods and give the position of the centers of the DC cap, but not their contours. The effective resolution of this method of observation is of about 3\mum, which imposes us to use solidification rates in the 0.01-0.1\mis range in the case of SCN-DC.

%%%%%%%%  {FRONTRODS}  %%%%%%%%%%
% 041405-1537
% 041405-1537VorGris
%%%%%%%%%%%%%%%%%%

%%%%%%%%  COURBURE %%%%%%%%%%
\begin{figure}[htbp]
\centering 
{\includegraphics*[width=8cm]{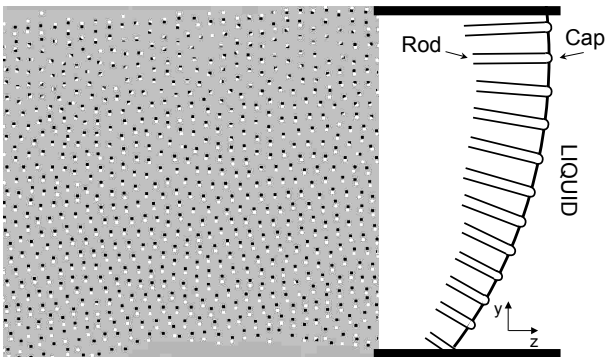}}
\caption{ {\it{Left}}: Superposition of two binarized micrographs of a growth pattern at times $t_1$ (black dots) and~$t_1+30$min (white dots). $V=0.035 \mu ms^{-1}$.  Vertical dimension: $400\mu m$.  {\it{Right}}: Sketch of a longitudinal section of the system assuming a curved growth front.}
\label{manip2}
\end{figure}
% front
%%%%%%%%%%%%%%%%%%

We registered the trajectories of all the DC caps appearing at the growth front during long solidification runs. This allowed us to reveal that, in all the experiments, the DC caps  were drifting  slowly  along the normal  to the sample walls  (Fig.  \ref{manip2}). No drift parallel to the walls was observed. The  drift velocity $V_d$ was a definite, roughly linear function of  $y$ that went to zero at a position $y_o$, which was slightly sample dependent, but was always located inside the sample. Thus the spatiotemporal diagram of the patterns was fan-shaped, and the patterns were  undergoing a permanent dilatation along {\bf y}. A simple explanation for this observation is that, in our setup,  the isotherm at $T_E$ has a negative curvature  (it is bulging toward the liquid) in the  plane $yz$ and that the trajectories of the DC caps remain locally perpendicular to this isotherm (Fig.  \ref{manip2}.b). While the latter statement, known as the normal-growth condition, is certainly correct to a good approximation, the former (a negative curvature of the isotherm) needs further discussion. 

% ($y_o$ could be changed by changing the thermal bias, confirming that the drift was linked to the thermal field)

The curvature of the isotherms in our system is slight, as will be seen shortly, and could not be measured by direct observation. We determined it from the drift of the growth pattern as follows. According to our conjecture, $V_d/V=\partial \zeta /\partial y$, where $z=\zeta(y)$ is the equation of the $T_E$ isotherm in the  plane $yz$. A linear fit on the measured values of  $V_d(y)$ gave us a linear function for $\partial \zeta /\partial y$, and, by integration, a parabolic function for $\zeta(y)$. TIt turned out that the radius of curvature $R$ at the vertex of the parabola ({\it{i.e.}} at $y=y_o$) was always much larger than the sample thickness. The curvature was thus independent of $y$ within the experimental uncertainty. Measurements of $R$  performed with this method in nine samples for different values of $V$ and  $\bar{\lambda}$  ranged from about $2.2$ to $5.4$mm. Incidentally,  an elementary calculation gives the residual thermal bias  knowing $R$, $y_o$ and $G$.   The maximum value of  $G_y$ was about $0.4\rm{Kmm^{-1}}$ in our experiments.

\begin{figure}[htbp]
\centering 
{\includegraphics*[width=5cm]{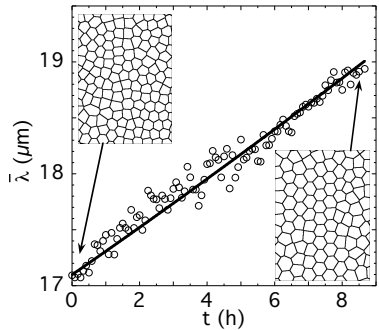}}
\caption{Time variation of the mean spacing in the absence of rod splitting events. $V=0.021  \mu ms^{-1}$. Open circles: measured $\bar{\lambda}$ values. Continuous line: exponential fit with a characteristic time of $82h$. Insets: Voronoi diagrams of a fixed region of the growth pattern at the beginning (upper left corner) and the end (lower right corner) of the process.}
\label{manip3}
\end{figure}

We obtained another estimate of $R$ from the rate of increase of $\bar{\lambda}$ over time. We define $\lambda$ and $\bar{\lambda}$ as the mean values of  the nearest-neighbor distances for a given DC cap and the whole growth front, respectively, at a given time $t$. An elementary geometrical calculation shows that, for an isotherm of uniform one-directional curvature $R^{-1}$, $\bar{\lambda}\approx \bar{\lambda}_0 e^{t/\tau}$, where  $\bar{\lambda}_0$ is the value of  $\bar{\lambda}$ at $t=0$ and $\tau=2R/V$. This equation is valid when the number of repeat units is locally conserved during the process, {\it{i.e.}} when no DC cap is created or eliminated. Whether or not this conservation condition was fulfilled during a given portion of a solidification run could be checked by direct observation.  DC cap elimination events were never observed (except, of course, at the glass walls) during this study, as could be expected since the pattern was under stretching, and can be left aside. Figure \ref{manip3} shows the time evolution of $\bar{\lambda}$ during a solidification run without creation events.  An exponential fit  (which was practically linear in the time interval of interest) yielded $\tau=82  \pm 12$h, and hence $R=3.2  \pm 0.5$mm, in good agreement with the value of $R$ derived from the measurement of  $V_d(y)$ during the same experiment.

What is the origin of the curvature of the isotherms in our directional-solidification setup?  Several factors could be at play, among which the differences in thermal conductivity of the various media composing the samples, the rejection of latent heat at the growth front, and convection-induced transverse gradients. However, only the first of these factors (non-uniform thermal conductivity) is likely to be important in our case because of the very low solidification rates used, and the order of magnitude of ratio of $R$ to the sample thickness (0.1) observed. In any case, it should be noted that  the effect of a slight curvature of the isotherm on growth patterning presumably reduces to the normal-growth condition, which is independent of the origin of the curvature.

%%%%%%%%  ETIREMENTS %%%%%%%%%%
%032306-1610M8-spmoy-insert
%%%%%%%%%%%%%%%%%%

\begin{figure}[htbp]
\centering 
{\includegraphics*[width=2cm]{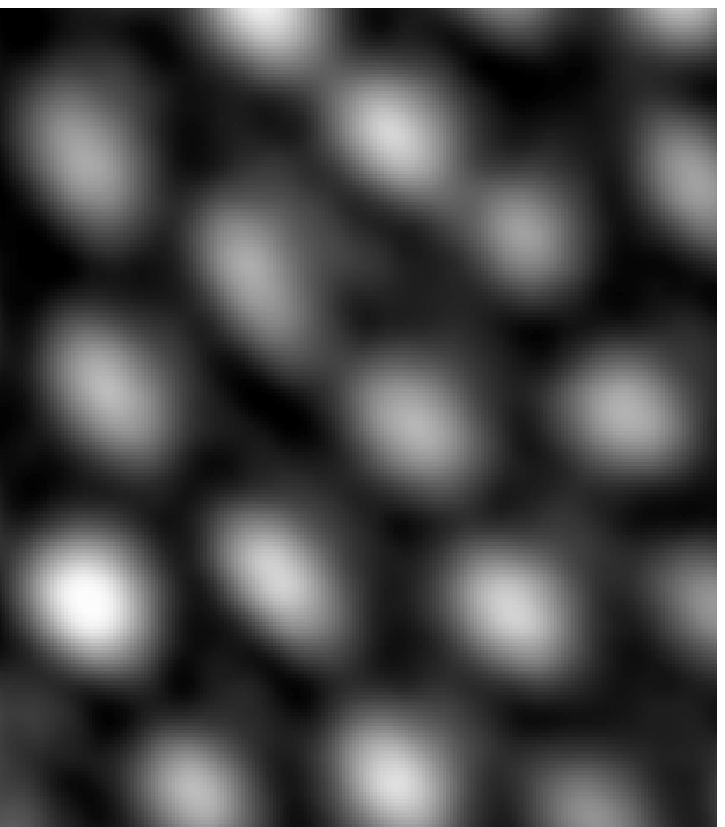}
\includegraphics*[width=2cm]{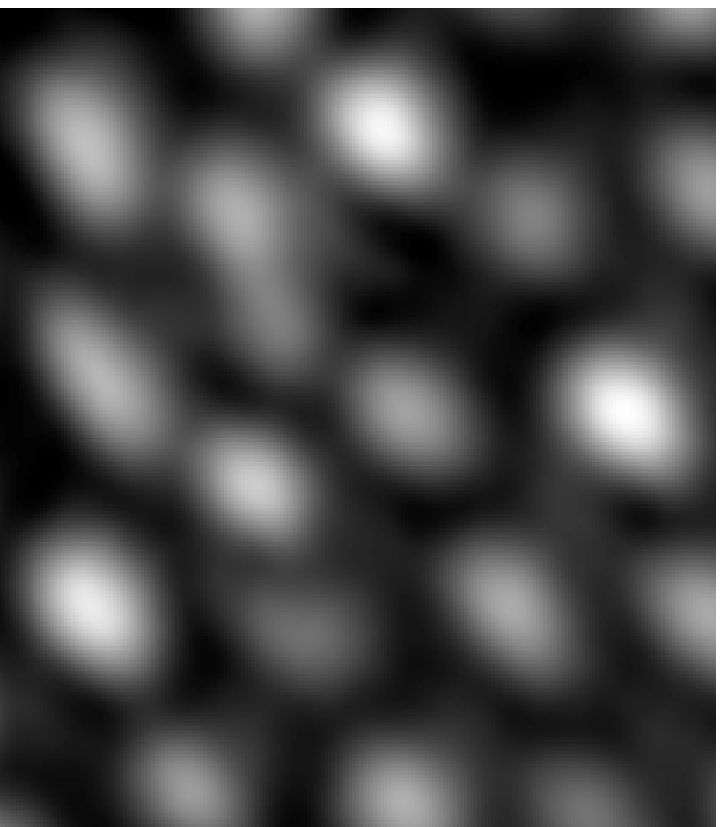}
\includegraphics*[width=2cm]{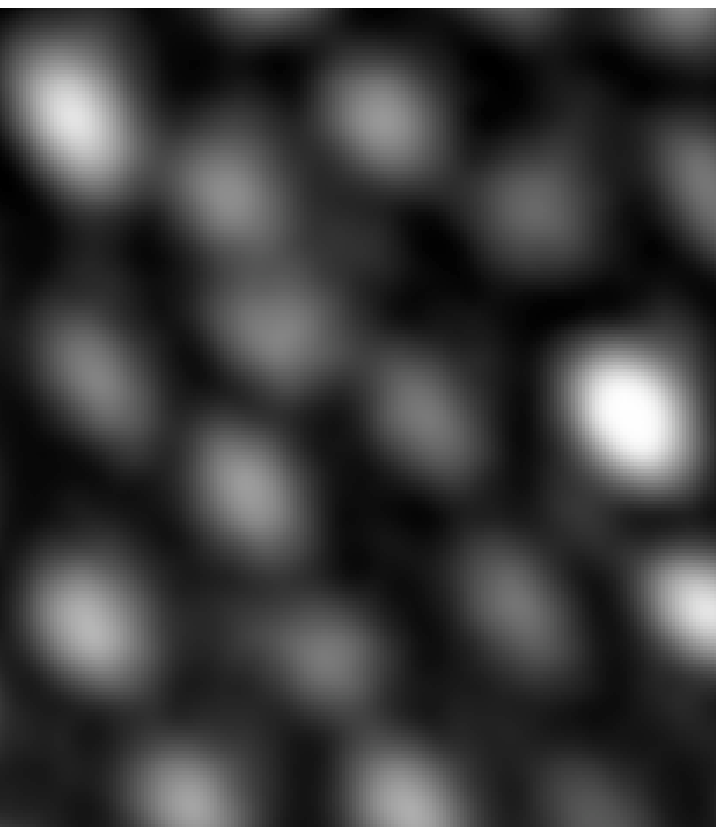}
\includegraphics*[width=2cm]{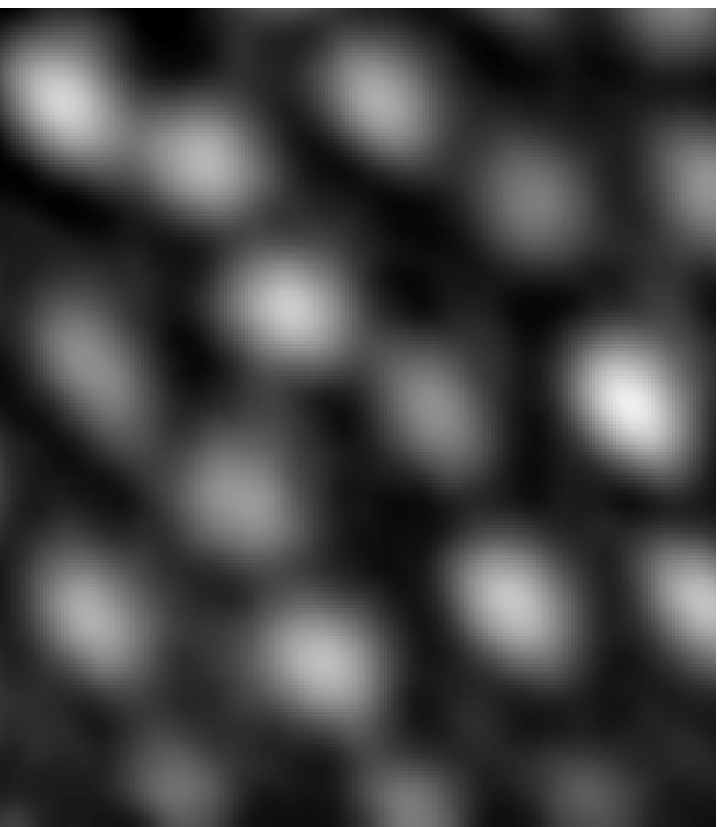}}
\caption{Direct observation of rod-splitting events. Snapshots of a $65\times74\mu m^2$ region of a growth pattern taken at time intervals of 7 min. $V=0.035 \mu ms^{-1}$. Two DC caps (located in the second column from left) are simultaneously  splitting. 
%The field of view has been translated in order to follow the global drift of the growth pattern
}
\label{manip4}
\end{figure}

\begin{figure}[htbp]
\centering 
{\includegraphics*[width=8cm]{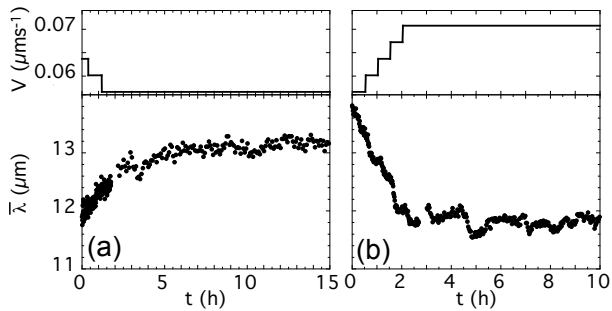}}
\caption{Time variation of the mean spacing $\bar{\lambda}$ during two different solidification rate programs. The time variations of $V$ and  $\bar{\lambda}$ are shown in the upper and lower parts of the diagrams, respectively. The initial value of the mean spacing  belonged to the stability interval in (a), and was above this interval in (b).  In the latter case, rod splitting occurred from the start, and became less and less frequent until a permanent regime was settled.}
\label{manip5}
\end{figure}

Bifurcations to oscillatory regimes apart, the  upper stability bound of  hexagonal eutectic patterns at fixed $V$   corresponds to an instability leading to the creation of new repeat units, and hence to a decrease in $\bar{\lambda}$ \cite{Perrut07}. This instability was observed in our experiments at sufficiently long solidification times. It consists of  the splitting of a DC cap into two parts  (Fig. \ref{manip4}). A detailed report on this, and other instability processes (rod termination, oscillatory bifurcations), will be presented in the future. For our present purpose, the important point is that, in long-duration experiments at constant $V$, the average  rod splitting frequency eventually adjusted itself so as to counterbalance the increase of   $\bar{\lambda}$ driven by the curvature of the isotherms. Any solidification rate program ending with a maintain at constant $V$ led to a permanent regime characterized by a plateau in the  $\bar{\lambda}(t)$ curve.  Illustrative examples are shown in Figure  \ref{manip5}. Direct counting of the rod splitting events confirmed that this plateau corresponded to a global balance between rod splitting and the  curvature-driven increase of  $\bar{\lambda}$.  In this regime,  $\bar{\lambda}$ fluctuated with characteristic times on the order of an hour about a constant value   $\lambda_p$.   Data from seven experiments performed with different solidification rate programs yielded  $\lambda_p \propto V^{-1/2}$  and $\lambda_p/\lambda_m=1.03  \pm 0.04$. A quantitative comparison between this quantity and  the threshold spacing for the rod splitting instability is beyond the scope of this article. However, it is worth noting that the local spacing values at the onset of rod splitting events were slightly larger than  $\lambda_p$.  

%%%%%%%%  {SPLITS}  %%%%%%%%%%
% split3
% split4
% split5
% split6
%%%%%%%%%%%%%%%%%%

\begin{figure}[htbp]
\centering 
{\includegraphics*[width=8cm]{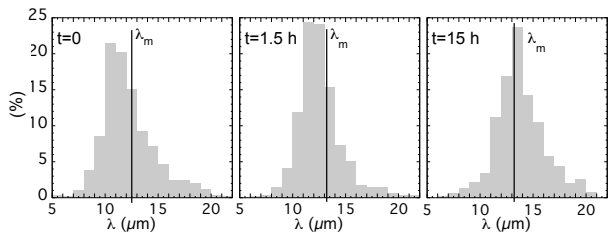}}
\caption{Normalized histogram of the local spacing at the indicated times during the solidification run of Fig.  \ref{manip5}.a.  $\lambda_m$: minimum-undercooling spacing }
\label{manip6}
\end{figure}

To sum up,  long-duration directional-solidification runs at constant $V$ comprised two stages: first, a transient, during which  the mean spacing $\bar{\lambda}$ increased or decreased depending on its initial value; second, a permanent regime, in which  $\bar{\lambda}$  fluctuated about a constant value  $\lambda_p$ slightly smaller than the threshold spacing for rod splitting.  What was the evolution of the degree of order of the growth patterns during these stages? During transients without rod splitting  (Fig.  \ref{manip5}.a), we observed a continuous ordering process, consisting in the formation and growth of perfectly ordered domains separated by thin boundaries (Figs.  \ref{manip1} and \ref{manip3}). Decelerating the solidification during such a transient made it last longer, which furthered the ordering process \cite{positiveCurv}.  The  local mechanisms of the ordering process remain to be studied. We think that the curvature-driven stretching of the pattern played an important role in this process, in particular, because the hexagonal array  had a nearest-neighbor direction either parallel or normal to the sample walls  inside  most domains. The order that was established during transients without rod splitting  was progressively destroyed when rod splitting set in. So long as there remained distinct domain boundaries, rod splitting events occurred preferentially in these boundaries, and had some short-range correlation.  The more or less periodic fluctuations of $\bar{\lambda}$ observed in the well-developed permanent regime are perhaps due to collective phenomena (such as the "cascades" observed in curved cellular growth fronts \cite{Bottin01}), but we  could not establish this clearly. For comparison with metallurgical studies,  the time evolution of the $\lambda$ distribution during a representative solidification run is shown in Figure \ref{manip6}. To note, the $\lambda$ distribution becomed peaked about $\lambda_m$, but still extended over a wide range in the permanent regime.  

%The transition between the two regimes corresponds to the onset of the latter instability.
%%%%%%%%  {LAMBDA DE T}  %%%%%%%%%%
% 031706-1618-1829LN
% 092905-1438-1758LN
%%%%%%%%%%%%%%%%%%

%%%%%%%%  {HISTOGRAMS}  %%%%%%%%%%
% distrib-lambda_567
%%%%%%%%%%%%%%%%%%

In conclusion, we have observed, in a transparent eutectic  directional-solidification system, a permanent regime, which presents statistical features (sharp selection of the mean spacing, broad dispersion of the local spacing values) similar to those that were previously noted, but not understood, in metallic  eutectics. We have brought to light the dynamical origin of this permanent  regime. A slight instrumental imperfection --namely, a negative curvature of the isotherms-- drives the system towards the marginal stability point for the repeat-unit creation instability of the growth pattern. After a transient, the two processes (external forcing and dynamical instability) come into balance, and the spacing remains close to the  instability threshold value. In general terms, this conclusion is not a surprise. Langer \cite{Langer1} pointed out long ago that the broad multistabilty of ideal directional-solidification systems  could be easily lifted by some external noise or forcing, and that, then, the system was likely to operate near a marginal stability point. Our study permits to specify that a curvature of the isotherms was the external forcing  at play in our system. A similar forcing is certainly at play  in all nonfaceted eutectic  directional-solidification systems, but it may be combined with other factors capable of producing a drift of the pattern such as capillary anisotropy and a transverse thermal bias. These factors were both small in our experiments, as mentioned, which allowed us to observe  the effect of  a curvature of the isotherms in their plainest form.

%%%%%%%%%%%%%%%%%%%%%%%%%%%%%%%%%%%%%%%%%%%%%%%%%%%%%%%%%%%%%%%%%%%%%%%%%%%%%%%%%%%%%%%%%%%%%%%%%%%%%%%%%%%%%%%%%%%%%%%%%%%%%%%%%%%%%%%%%%%%%%%

\section{Acknowledgments}
We thank  V.T. Witusiewicz, L. Sturz, and S. Rex from ACCESS (Aachen, Germany) for kindly providing us with purified  Succinonitrile-(d)Camphor  alloy.  This work was supported by the Centre National d'Etudes Spatiales, France.


\begin{thebibliography}{99}
\bibitem{chapeau} For detailed statistical data on the permanent regimes of metallic lamellar eutectics, see ref. \cite{TriMas91}. For information on rod-like eutectic growth patterns, see refs. \cite{Drevet96,Ratke00}.  These studies, unlike ours,  were performed in samples containing several eutectic grains.
\end{thebibliography}
\end{document}